# Biomimetic hierarchical structuring of PLA by ultra-short laser pulses for processing of tissue engineered matrices: study of cellular and antibacterial behavior.

Albena Daskalova[1*], Liliya Angelova[1], Emil Filipov[1], Dante Aceti[1], Rosica Mincheva[2], Xavier Carrete[2], Halima Kerdjoudj[3], Marie Dubus[3], Julie Chevrier[3], Anton Trifonov[4] and Ivan Buchvarov[4]

[1] Laboratory of Micro and Nano-Photonics, Institute of Electronics, Bulgarian Academy of Sciences, Sofia, Bulgaria; lily1986@abv.bg; albdaskalova@gmail.com; emil.filipov95@gmail.com; acetidm@gmail.com
[2] Laboratory of Polymeric and Composite Materials (LPCM), Center of Innovation and Research in Materials and Polymers (CIRMAP), University of Mons, Mons, Belgium; Rosica.MINCHEVA@umons.ac.be; Xavier.CARETTE@umons.ac.be
[3] Université de Reims Champagne Ardenne, EA 4691, Bomatériaux et Inflammation en site osseux BIOS, Reims, France; halima.kerdjoudj@univ-reims.fr; marie.dubus@hotmail.com; julie.e.chevrier@orange.fr
[4] Faculty of Physics, St. Kliment Ohridski University of Sofia, Bulgaria; a.trifonov@phys.uni-sofia.bg; ivan.buchvarov@phys.uni-sofia.bg
* Correspondence: albdaskalova@gmail.com;

**Abstract:** The influence of ultra-short laser modification on the surface morphology and possible chemical alteration of poly-lactic acid (PLA) matrix in respect to optimization of cellular and antibacterial behavior were investigated in this study. Scanning electron microscopy (SEM) morphological examination of the processed PLA surface, showed the formation of diverse hierarchical surface microstructures, generated by irradiation with a range of laser fluences (F) and scanning velocities (V) values. By controlling the laser parameters, diverse surface roughness can be achieved, thus influencing cellular dynamics. Such surface feedback can be applied to finely tune and control diverse biomaterial surface properties like wettability, reflectivity, and biomimetics. The triggering of thermal effects, leading to the ejection of material with subsequent solidification and formation of raised rims and 3D-like hollow structures along the processed zones, demonstrated a direct correlation to the wettability of the PLA. A transition from superhydrophobic ($\theta >150°$) to super hydrophilic ($\theta <20°$) surfaces can be achieved by the creation of grooves with V=0.6 mm/s, F= 1.7 J/cm². The achieved hierarchical architecture affected morphology and thickness of the processed samples which were linked to the nature of ultra-short laser-material interaction effects, namely the precipitation of temperature distribution during material processing can be strongly minimized with ultrashort pulses leading to non-thermal and spatially localized effects that can facilitate volume ablation without collateral thermal damage The obtained modification zones were analyzed employing Fourier transform infrared (FTIR), X-ray photoelectron spectroscopy (XPS), Energy dispersive X-ray analysis (EDX) and optical profilometer. The modification of the PLA surface resulted in an increased roughness value for treatment with lower velocities (V=0.6 mm/s). Thus, the substrate gains a 3D-like architecture and forms a natural matrix by microprocessing with V=0.6 mm/s, F=1.7 J/cm², and V=3.8mm/s, F= 0.8 J/cm². The tests performed with Mesenchymal stem cells (MSCs) demonstrated that the ultra-short laser surface modification altered the cell orientation and promoted cell growth. The topographical design was tested also for the effectiveness of bacterial attachment concerning chosen parameters for the creation of an array with defined geometrical patterns.

## 1. Introduction

According to data from World Population Prospects: the 2019 Revision, (reference date: 01.07 2020), 24.2% of all the people in the world are over the age of 50 [1]. This fact determines the recrudescence of diseases such as bone fragilities, osteoarthritis, cartilage, and cardiovascular pathologies as one of the major problems of modern societies. Tissue engineering, as an innovative field in biotechnology, deals with the creation, regeneration, and improvement of the function of injured or missing tissues. The design of smart porous biocompatible and biodegradable scaffolds, in association with specific tissue cells and growth factors, plays a crucial role in the regeneration of tissues and recovery of their functionality [2]. The main purpose of the engineered scaffold is to create a biomimetic environment that stimulates cell adhesion, proliferation, and differentiation; as a result, the ingrowing cells are reorganized into new three-dimensional tissue [3]. During this gradual biodegradation of the scaffold, a new tissue with the desired properties is naturally formed. The selection of appropriate biocompatible source materials is crucial for successful scaffold fabrication [4, 5]. Polylactic acid (PLA) has extremely good mechanical properties (stability, toughness, elasticity, and strength), biocompatibility, biodegradability, ease of processing, and thermal stability [6-8]. The up-listed properties are the main features for the successful application of PLA-engineered constructs in the regeneration of tissues, undergoing mechanical stress under normal physiological conditions [3, 9]. PLA is an aliphatic polyester with backbone formula $(C_3H_4O_2)_n$ or $[-C(CH_3)HC(=O)O-]_n$, approved by the US Food and Drug Administration (FDA) for different biomedical applications, [10] for which the synthetic biopolymer is in direct contact with biological fluids [6,11]. An important advantage of this polymer is that it is naturally resorbed as the physiological repair process takes over the main product of PLA metabolism in mammals is lactic acid [12]. It is incorporated in the carboxylic acid cycle and completely excreted by natural degradation pathways [13]. The biocompatibility of the scaffold generally refers to inflammatory response, as well as to the intrinsic micro and nanoscale structure of the biosurfaces. On the other hand, PLA is a hydrophobic polymer, which lowers its bioactivity, regarding the interaction of the scaffolds with cells, tissue, and fluids [14-22]. Additional surface functionalization is needed for the creation of effective bio-interfaces between the cells seeded and the PLA-based scaffold [23]. This could be achieved by different chemical methods, most of which, however, use organic compounds in order to control surface properties of the tissue scaffolds (by modifying surface chemistry), leaving residual toxic components in the cells environment and thus may lead to risks of cells toxicity and carcinogenicity [24]. As an alternative, the properties of various cell matrices could also be controlled by the means of different physical methods (by controlled removal, addition, or deformation of the biomaterial surface in order to increase its roughness) [25, 26]. The most innovative and bio-friendly approach are the laser-based techniques due to their non-contact nature of interaction with the material and thus the absence of chemicals used for additional treatment of the created cell/tissue matrices. In the literature, there is scarce information concerning femtosecond laser modification of PLA. Irradiation with ultra-short laser pulses has been extensively studied in the last years, especially for the functionalization of tissue-engineered scaffolds, based on various biomaterials [27, 28]. Femtosecond (Fs) laser-based processing is an alternative approach, characterized with extremely high processing accuracy and absence of thermal deformations on the processed material due to its extremely high-peak power and pulse duration which is shorter than material thermal relaxation time. Ultra-short pulses lead to non-thermal and spatially localized effects, a result of which volume ablation occurs without collateral thermal damage to the surrounding zones. The effects of temperature distribution during the interaction with extremely short pulses is strongly minimized. The creation of micro and nanostructures on the surface of the biomaterial can strongly affect cell behavior, as diverse surface roughness (in macro, micro, and nanoscale) can be achieved by controlling the applied laser parameters [29]. Such surface feedback can be used to finely tune and control diverse properties of the processed material like wettability and biomimetics. Wettability is an important factor in cell adhesion, and the fs laser surface modification has the ability to aid in tuning it in respect to the effect of interest - namely achieving cell adhesive/antiadhesive scaffold surfaces, needed for diverse tissue regenerative applications [30-33] or even creation of enhanced antibacterial cell surface environment [34]. Yada and Terakawa even achieved the generation of laser-induced periodic surface structures (LIPSS) on a highly-biocompatible poly-L-lactic acid scaffold by femtosecond laser processing [35]. These structures are examined for cell behavior control. In the current study, PLA polymer surfaces were modified by Ti: sapphire femtosecond laser at diverse laser energies and scanning velocities were applied. Different surface treatments were used to obtain structures with hierarchical geometries for future preparation and design of porous PLA-based cellularized scaffolds and bio-interfaces between the tissues of the recipient and the foreign implant. Laser microstructured PLA scaffolds were investigated by SEM, EDX, XPS, optical profilometer studies, wettability and FTIR. The thickness of the samples, before and after ultra-fast laser modification, was compared. *In vitro* degradation tests and preliminary cellular studies were accomplished. Mesenchymal stem cells from Wharton Jelly (MSCs) were cultured on PLA scaffolds for evaluating the effect of the patterned polymer surface on cell proliferation and cell morphology from day 1 to 14, compared to control PLA matrix. Adhesion of *Staphylococcus aureus* was also evaluated for the potential antibacterial effect which laser treatment could achieve. Thus, the aim of this examination will be directed towards achievement of conditions, associated with regime above modification threshold, and which gives a feedback concerning to which limit the PLA material could be processed without triggering chemical state alterations. This would potentially lead to optimization of laser patterning conditions in order to achieve desired improved morphologies for cellular adhesion and bacterial rejection. The results obtained demonstrate that the precise control of the laser parameters applied to the PLA scaffolds could generate bioactive surfaces by means of the nondestructive Fs laser modification technique.

**2. Materials and Methods**

*2.1. Preparation of PLA samples*

Polylactic acid (PLA) in the form of amorphous compound was purchased from Nature Works, Nebraska, USA (PLA 4060D). Compression molding was used for the PLA samples preparation - the thermos compression of the PLA pellets was performed by Carver 4122 12-12H Manual Heated Press (Carver Inc., USA). First the pellets were vacuum dried overnight at 60°C (Tg PLA=60°C), then inserted in the mold (square of 0.5mm thick). The procedure used for the compression consists of the following steps: 3 min of contact at 180°C, several cycles of degassing, and 2 min at 12 bars. The as prepared PLA was cut into squares of 2x2cm.

*2.2. Fs laser experimental Setup*

The synthesized PLA samples were laser processed by Quantronix- Integra-C Ti: sapphire (Hamden, CT, USA) Fs laser system ($\tau$ =150fs) at a central wavelength of $\lambda$ = 800 nm and 0.5 kHz repetition rate Figure1. For samples positioning perpendicular to the laser beam, a high-precision XYZ translation stage was used. All the experiments were performed in the air by scanning the laser beam over the material surface at precisely defined separation intervals (in the X and Z axis) in order to optimize the PLA surface texturing. All modifications of the PLA plates, performed by the Integra-C laser system were precisely controlled by LabView software. On the basis of previously performed large scale preliminary experiments on PLA, part of which are already published [36], the following laser parameters were chosen for surface structuring of the material: fluence F=1.7 and 0.8 J/cm² and scanning velocity V=16, 3.8, 1.7, and 0.6 mm/s for both F values used. Each laser processed PLA scaffold was analyzed with respect to the control surface.

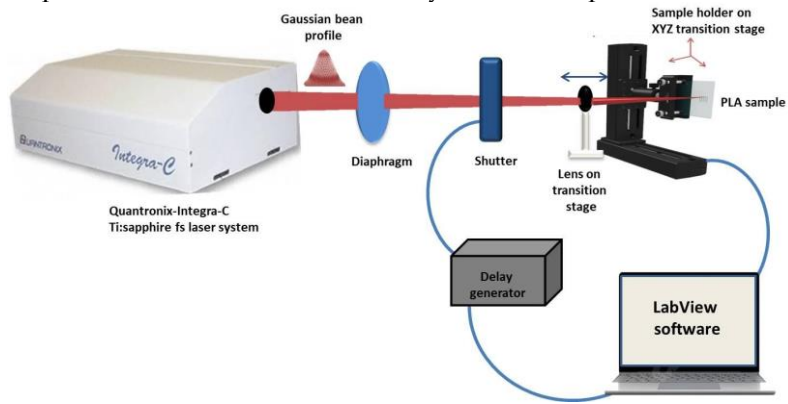

**Figure 1.** A scheme of the optical setup for Fs laser patterning of PLA sample positioned perpendicular to the laser beam on XYZ translation stage.

*2.3. Laser-modified samples qualitative characterization methods*

2.3.1. Thickness measurement

For determination of the thickness of the PLA samples before and after laser treatment, a coating thickness gauge VA 8042 coating meter (Zhejiang, China) was used, and every value presented (in μm) is averaged over 10 separate measurements.

2.3.2. Roughness measurement

For evaluation of the surface topography of the PLA samples, before and after fs laser treatment, 2D and 3D roughness analysis was performed via 3D Optical profiler, Zeta-20 (Zeta Instruments, KLA, Milpitas, California). It is a fully integrated microscope (magnifications from 5x to 100x) that provides 3D imaging as it scans a sample over specified vertical range. The Zeta Optics Module of the profilometer records the XY location and the precise Z height at each Z position scanned (vertical (Z) resolution < 1 nm). As a result true color 3D and 2D composite images are created and roughness parameters $R_a$ (arithmetical mean height of a line - the mean value of the deviations of the surface height from the median line, according to DIN4776 standards) and $S_a$ (the extension of $R_a$ to a surface) are obtained. Every value for $R_a$ and $S_a$ is averaged over 5 roughness measurements of every Fs-processed and control PLA sample surface area. For better visualization and data processing ProfilmOnline software (https://www.profilmonline.com) was additionally used.

2.3.3. WCA evaluation

The water contact angle measurement measurements were performed in air with distilled water drop with volume of 1μL on control and laser modified surfaces of the PLA samples for the time period of 0.5-7s - Figure 2. The samples were positioned on XYZ horizontal translation stage. The measurements were performed according to a contact angle goniometry method - the droplet was deposited by a fixed above the sample micropipette (Dlab, Beijing, China ISO9001/13485 certified -0.1-10 μL). The process was captured by a high resolution camera (Huawei P40 pro camera in super macro mode with additional macro lens - Huawei Investment & Holding Co., Ltd., Shenzhen, China). For WCA evaluation, the images obtained were analyzed by *ImageJ* software equipped with contact angle measurement plug-in. Measurements were performed independently in two directions - along and perpendicular to the Fs grooves in respect to control PLA surface for a period of 7 seconds. Every WCA value presented is averaged over 10 separate measurements.

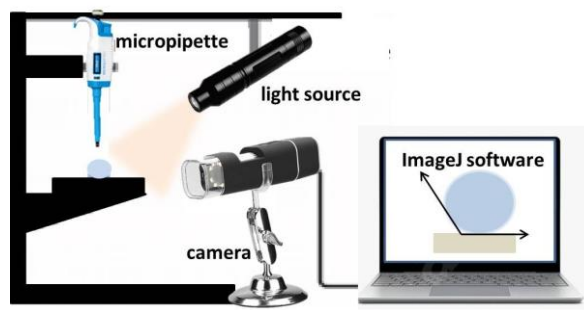

**Figure 2.** Experimental setup for measuring the water contact angle (WCA) of PLA samples.

2.3.4. SEM-EDX analysis

The surface morphology and elemental composition, before and after Fs laser treatment were analyzed by Scanning Electron Microscopy (SEM) Hitachi SU8020 (Hitachi High-Technologies Corporation, Tokyo, Japan), with field emission gun with landing energy at 5 kV and SE (UL) detector. The samples were sputter- coated with a nm layer of gold (Au) and several images for every modification were taken at magnification 100x, 500x, 5000x. Energy Dispersive X-Ray analysis (EDX) was also performed - the corresponding elemental composition [wt. %] was defined.

2.3.5. FTIR analysis

Fourier-Transform Infrared (FTIR) spectrophotometer (IR Affinity-1, Shimadzu, Kyoto, Japan), with resolution of 4 cm$^{-1}$ and working range 4500 - 500 cm$^{-1}$, was used for obtaining the IR transmittance spectra [%] of the Fs treated PLA plates for evaluation of possible chemical bonds alterations after laser treatment. Every measurement was related to a control PLA spectrum [37].

2.3.6. XPS analysis

AXIS Supra electron spectrometer (Kratos Analytical Ltd., Manchester, UK) was used for X-ray photoelectron spectroscopy (XPS) measurements of control and laser processed PLA samples. High-resolution detailed spectra of $C_{1s}$ and $O_{1s}$ (%) were taken for identifying the elemental composition and chemical state of the PLA samples of interest (laser treated with F=1.7 and 0.8 J /cm$^2$, V=0.6 and 3.8mm/s, respectively), according to control surface.

2.3.7. *In vitro* degradation test

A selected group of fs laser modified PLA samples, as well as control samples were subjected to *in vitro* degradation test – the samples were stored in an incubator at 37$^0$C for 8 weeks in PBS buffer saline (pH 7.2, Sigma-Aldrich, Missouri, United States). Each sample was put in 3ml PBS, the pH of the solution was checked every week with a pH meter (VAT1011, V & A Instrument Co.,Ltd., Shanghai, China) with an external sensor, and the PBS solution was replaced with a fresh buffer.

*2.4. Cell behavior on laser-textured PLA matrices*

Two groups of Fs treated samples were chosen for preliminary cellular experiments – **Group 1 (G1)** – PLA processed with F= 0.8 J/cm$^2$, V=3.8mm/s and **Group 2 (G2)** – PLA processed with F= 1.7 J/cm$^2$, V=0.6mm/s. Each group consisted of 30 samples; in order to confirm the obtained results the performed cell seeding experiments were repeated twice, as parallel cell seeding was performed on control (laser untreated PLA samples – **Group 3**

**(G3)**). PLA samples from the three groups were cleaned with 70% ethanol for 15min in a 12-well plate Figure 3.

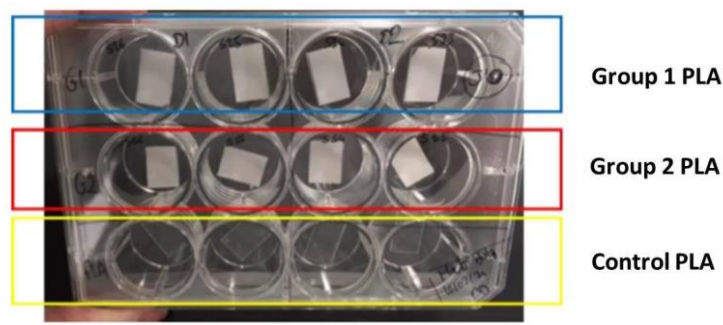

**Figure 3.** Preparation procedure of cell culture studies.

The plates were placed under a laminar flow hood to help the remaining ethanol to evaporate and then sterilized under a UV lamp, for 20 min, before any other manipulation. Mesenchymal stem cells from Wharton Jelly (MSCs) were seeded at a density of 30 000 cells per sample and were cultured for 14 days. WST-1 cell proliferation assay (Roche Diagnostics, Risch-Rotkreuz, Switzerland) was performed at day 7, day 10 and day 14 on MSCs in accordance with the manufacturer protocol – Figure 4. Absorbance was measured at 440 nm using a FLUOstar Omega microplate reader (BMG Labtech, Aylesbury, England) against a background control as blank. A wavelength of 750 nm was used as the correction wavelength.

After 14 days of culture, MSCs were fixed with 4% (w/v) paraformaldehyde (Sigma-Aldrich, Missouri, United States) at 37 °C for 10 min and permeabilized with 0.5% (v/v) Triton ×100 for 5 min. Alexa® Fluor-488 conjugated-Phalloidin® (Abcam, Cambridge, United Kingdom) - 1/100 dilution in 0.1% Triton ×100 was used to stain F-actin cytoskeleton for 45 min at room temperature. Nuclei were counter-stained with 4,6-diamidino-2-phenylindole (DAPI, 100 ng/mL, 1/10 000 dilution) for 5 min. Stained cells were mounted and imaged by epifluorescence microscopy (Zeiss microscope, ×20, Carl Zeiss AG, Jena, Germany).

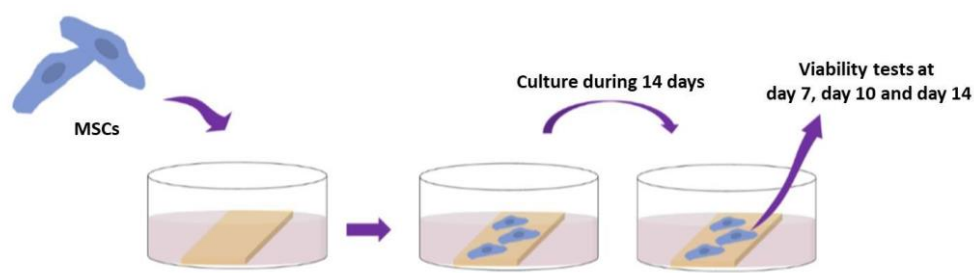

**Figure 4.** Cell viability test performed at day 7, day 10 and day 14.

*2.5. Microbiology studies*

Duplicate PLA samples from the second group (**G2**) - processed with F= 1.7 J/cm², V=0.6 mm/s were chosen for performing antibacterial studies. Parallel seeding was performed on control, laser untreated PLA samples – Group 3 (**G3**). Both groups consisted of 12 samples; experiments were performed twice in order to confirm the obtained results. All samples were incubated with 70% ethanol for 15min in a 12-well plate. Remaining ethanol was thrown out, and plates were placed under a laminar flow hood to help ethanol evaporate. Plates were then sterilized under a UV lamp for 20 minutes. Adhesion of *Staphylococcus aureus* (SH1000 strain), commonly found in nosocomial infections, was evaluated by incubating the PLA scaffolds for 24h with a *S. aureus* suspension. Then *S.*

*aureus* adhered on **G2** and **G3** PLA samples were enumerated to obtain the number of adhered bacteria on each sample. Three independent experiments were carried out, on samples from each group, by the protocol presented on Figure 5.

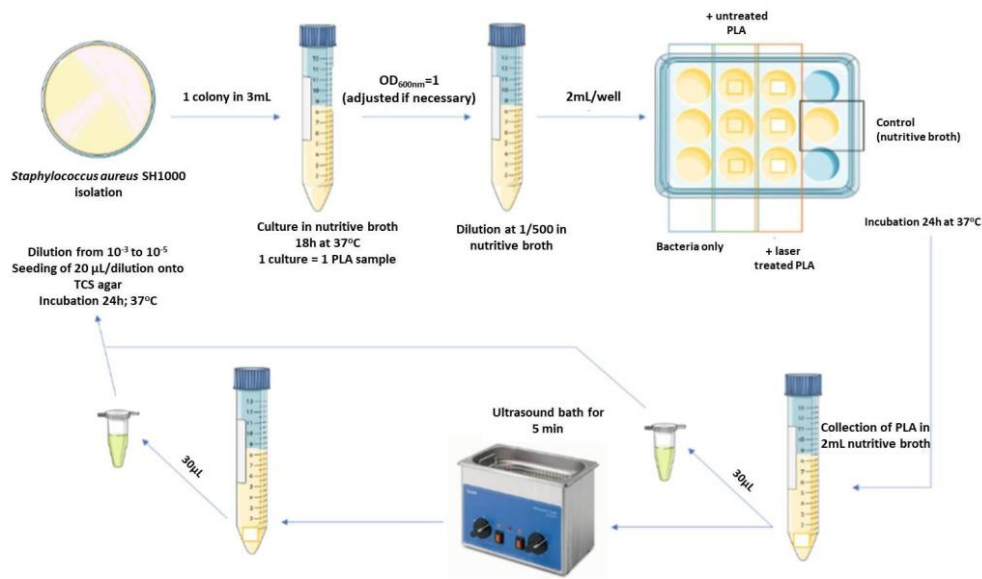

**Figure 5.** Scheme of the used microbiology protocol.

## 3. Results and Discussion

### 3.1. Morphological, structural and chemical characterization of Fs modified PLA specimens

After synthesis and Fs modification of PLA samples, SEM analysis was performed for evaluating the surface morphological change, according to applied laser parameters, chosen for surface structuring of the samples: fluence $F=0.831 J/cm^2$ and scanning velocity $V=16, 3.8, 1.72$ and $0.551 mm/s$ (Figure 6 a-d, respectively) and $F=1.663 J/cm^2$ and $V=16, 3.8, 1.72$ and $0.551 mm/s$ (Figure 6 e-h, respectively). Each Fs modification of the PLA scaffolds is observed at 100x, 500x and 5000x magnification. The corresponding values of C [wt. %] and O [wt.%] of the performed EDX elemental composition analysis are given in Table 1.

The surface of the control PLA matrix has a value of $S_a=0.41 \mu m$, compared to the laser treated samples where change of the $S_a$ values in the range of $0.81 \div 11.84 \mu m$ was monitored – Table 2. By varying the laser parameters, very precise control over surface and even volume structuring could be achieved – from very gentle surface round formations at $V=16 mm/s$, $F=0.8 J/cm^2$, through holes formation and ejection of material around the zones of interaction and formation of grooves with different depths and widths (Table 2) for $V=3.8, 1.7, 0.6 mm/s$, $F=0.8 J/cm^2$ and $V= 16, 3.8, 1.7 mm/s$, $F=1.7 J/cm^2$. It was observed creation of self-formed complex frameworks with two types of periodicities – parallel and perpendicular to the direction of the laser processing is achieved at $V=0.6 mm/s$, $F=1.7 J/cm^2$ - Figure 6 (h). The interaction with the highest values of V ($V=16 mm/s$, for both values of F) with the PLA surface leads to formation of single craters saturated with porous structures Figure 6 (a) and (e). By further increase of applied laser fluence and lowering the scanning velocity transformation into the groove-like structure is observed with increment in depth and width. Thus a 3D foam-like porous structure is formed on the treated PLA surface. These hierarchical forms represent a very fine network of filaments, which is clearly seen in the images with the highest magnification ($V=16, 3.8 mm/s$ at both F - Figure 6 (b-c) and (f-g)). In the case of $V=0.6 mm/s$, $F=1.7 J/cm^2$, Figure 6 (h) two types of perpendicular to each other periodical structures are observed. Such laser-induced surface structures are formed due to the elevation of the material and are known to be very effective for controlling cell behavior.

In addition to that, several groups state that these structures could lead to an actual reduction in bacterial biofilm formation and thus contribute to the creation of scaffolds with enhanced antibacterial properties [34].

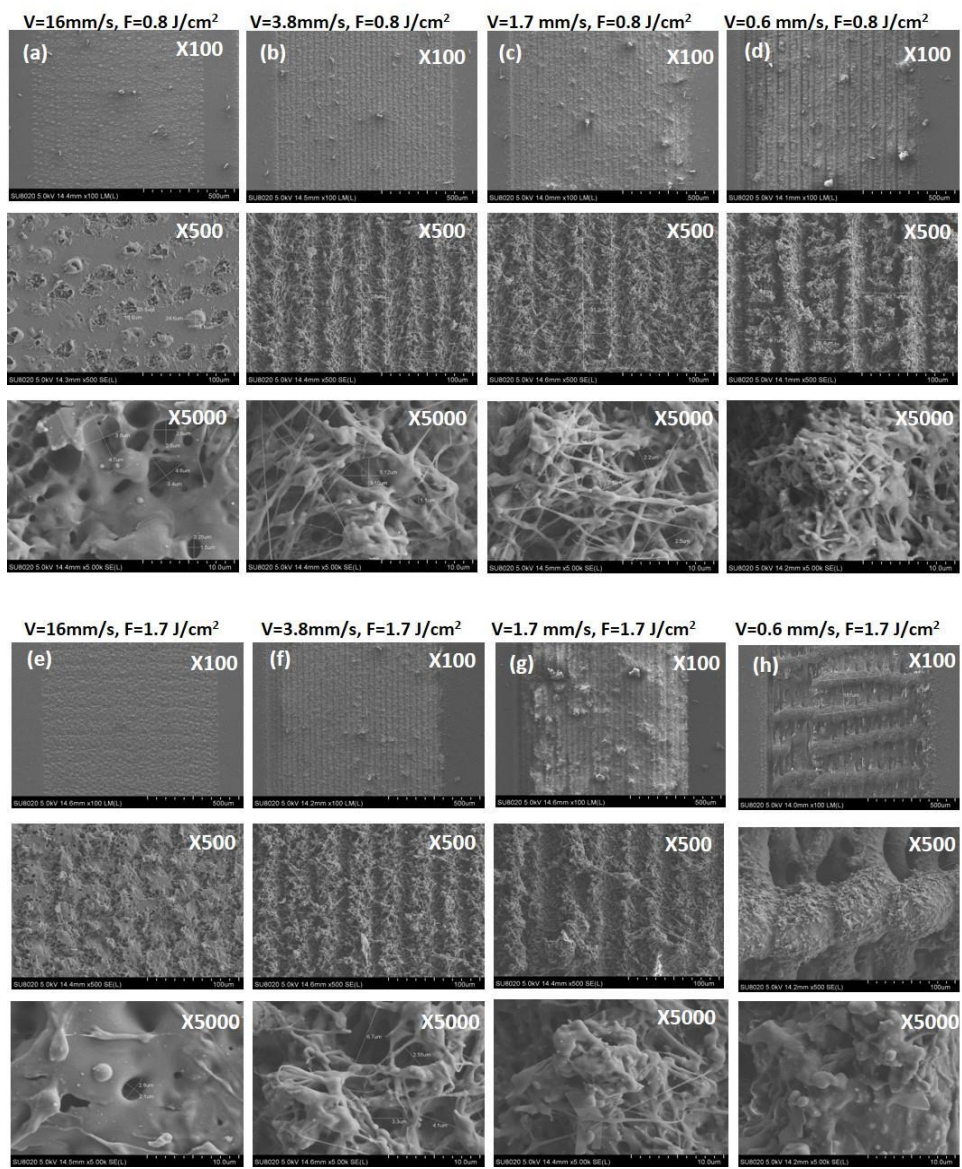

**Figure 6.** SEM images of PLA matrices: **(a-d)** patterned with F=0.8 J/cm$^2$ and V=16, 3.8, 1.7, 0.6 mm/s, respectively **(e-h)** and F=1.7 J/cm$^2$ and V=16, 3.8, 1.7, 0.6 mm/s respectively.

EDX elemental composition [wt. %] values obtained from the laser processed and from control surfaces of PLA matrices are presented in Table 1. According to the EDX results, the presence of uncommon elements in the composition of all PLA samples was not detected. The breakage of side O-C=O chemical bonds could explain the slight variation in the percent concentration [wt.%] of oxygen [O] and carbon [C] observed with increasing values of laser fluence (F) and lowering the scanning velocity (V).

**Table 1.** EDX elemental composition values [wt. %] of the control and Fs laser modified PLA samples.

| Sample (PLA) | C [wt.%] | O [wt.%] |
| --- | --- | --- |
| Control | 74.27 | 25.73 |
| V=16 mm/s, F=0.8 J/cm$^2$ | 68.05 | 31.95 |
| V=3.8 mm/s, F=0.8 J/cm² | 61.51 | 38.49 |
| V=1.7 mm/s, F=0.8 J/cm² | 60.52 | 39.48 |
| V=0.6 mm/s, F=0.8 J/cm² | 58.82 | 41.18 |
| V=16 mm/s, F=1.7 J/cm² | 62.48 | 37.52 |
| V=3.8 mm/s, F=1.7 J/cm² | 60.42 | 39.58 |
| V=1.7 mm/s, F=1.7 J/cm² | 59.09 | 40.91 |
| V=0.6 mm/s, F=1.7 J/cm² | 58.32 | 41.68 |

In parallel with these measurements thickness and roughness evaluation of the Fs processed PLA samples were performed. The results are summarized in Table 2. Each test was done in relation to the control surface.

**Table 2.** Thickness [μm] and Roughness (**Mean average roughness (**$R_a$) and **Areal surface roughness** ($S_a$) [μm] ) of the laser processed PLA samples, compared to control surface

| Sample (PLA) | Thickness [μm] | $R_a$ [μm] | Width [μm] | Depth [μm] | $S_a$ [μm] |
|---|---|---|---|---|---|
| Control | 250 | 0.024 | - | - | 0.41 |
| V=16 mm/s, F=0.8 J/cm² | 228 | 0.53 | 0.23 | 2.85 | 1.38 |
| V=3.8 mm/s, F=0.8 J/cm² | 215 | 1.92 | 1.19 | 6.30 | 2.10 |
| V=1.7 mm/s, F=0.8 J/cm² | 219 | 2.4 | 5.67 | 15.23 | 3.64 |
| V=0.6 mm/s, F=0.8 J/cm² | 236 | 4.24 | 36.01 | 50.42 | 11.84 |
| V=16 mm/s, F=1.7 J/cm² | 227 | 3.35 | 0.28 | 2.88 | 0.81 |
| V=3.8 mm/s, F=1.7 J/cm² | 223 | 5.8 | 2.91 | 27.75 | 3.25 |
| V=1.7 mm/s, F=1.7 J/cm² | 218 | 4.0 | 2.91 | 43.22 | 9.72 |
| V=0.6 mm/s, F=1.7 J/cm² | 240 | 5.19 | 9.67 | 61.87 | 24.07 |

The results presented in Table 2 demonstrate a tendency of thinning the samples with increasing laser fluence (F) and lowering the scanning velocity (V) applied to the sample's surface. Laser treatment with V=0.6 mm/s at F=0.8 J/cm2 and F=1.7 J/cm2 represents a turning point since it leads to elevation of the material around the laser treated zones, creating a local 3D structuring on the surface of the PLA plate. These findings could be explained by bumps formation from elevated material in the high energy regime of processing, which could be clearly seen on the SEM images Figure 6 (d) and (h). This effect could be attributed to the formation of melting and severe ablation of the solid PLA matrix leading to the formation of a structure with a hierarchy of repeating depths and heights. The monitored change could be explained by rapid increase in surface temperature, followed by sudden partial melting of the polymer matrix and confinement of phase transformation. This is affecting the thickness of the processed region in the case of the lowest values of scanning velocity (V) at F=0.8 J/cm² and F=1.7 J/cm². The high roughness values ($S_a$=11.84μm and 24.07μm) achieved at the above mentioned laser processing conditions, could not have a positive effect only on cells attachment and guidance, but also on the orientation of the cellular ingrowth towards zones with a greater porosity in volume as provided by the two types of periodicities achieved at V=0.6 mm/s, F=1.7 J/cm². On the other hand, the surface roughness ($S_a$=2.10μm) achieved at F=0.8 J/cm² at V=3.8 mm/s, in respect to depth (6.30μm) and width (1.19μm) of the grooves, as well as to the observed morphology at the SEM images - Figure 6 (b) and the $R_a$=1,92μm in this specific case seems appropriate for mesenchymal stromal cells (MSCs) seeding, according to Szmukler-Moncler et al., who reported $R_a$ = 0.9 - 1.53 μm as optimal for MSCs [38]. The optimum roughness for stable bonding of the implant surface to recipient tissue, according to literature, is min. $R_a$=1μm [39, 40].

For better visualization of the detailed results, presented in Table 2, representative 3D reconstructed images of the control and processed PLA matrices are shown in Figure 7.

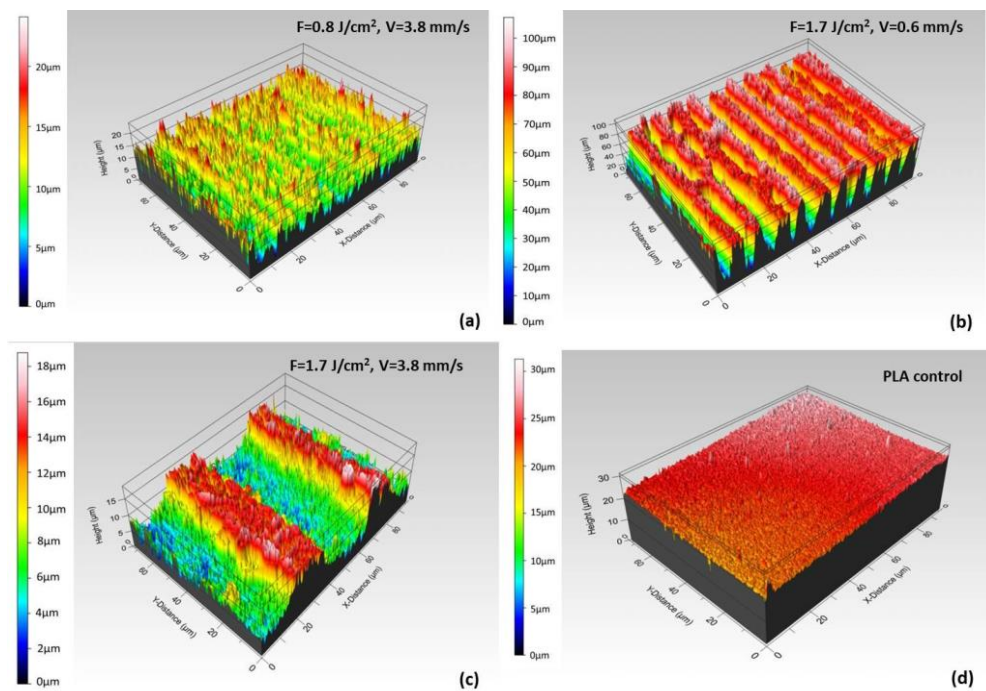

**Figure 7.** Representative 3D cross section images of PLA matrices: **(a)** F=0.8 J/cm², V=3.8mm/s, **(b)** F=1.7 J/cm², V=0.6 mm/s, **(c)** F=1.7 J/cm², V=3.8mm/s; **(d)** control surface.

The WCA measured on structured surfaces depends on the direction of water drop application in respect to the grooves direction – Figure 8. Every WCA value presented on Figure 9 graphs is averaged over 10 separate measurements.

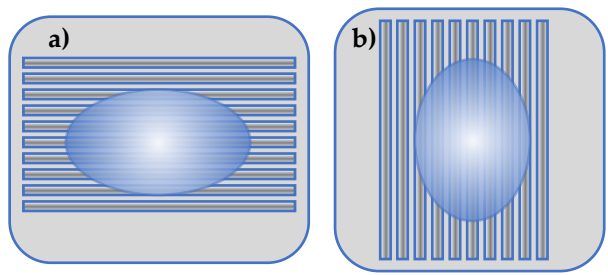

**Figure 8.** Scheme of water droplet positioning of Fs modified PLA samples: **(a)** perpendicular to the laser created grooves and **(b)** along the laser created grooves;

As can be clearly seen from the graphically presented results in Figure 9, in both types of droplet application, a general trend of lowering the water contact angle values is observed by increasing the contact time of the water drop with the surface at higher values of applied laser fluence and lower scanning velocity. The monitored change of WCA ($\theta$) values from 137° for control surface to 60°-20° for processed PLA transfers the surface wettability properties to more hydrophilic ones, thus making it "cell adhesion and proliferation friendly", which could improve the tissue integration of PLA implants [41]. The groups of Dekker et al. and Wan et al. for example, report that maximal cells attachment strength occurs at WCA ($\theta$) of 20–55° for endothelial cells and fibroblasts [42, 43].

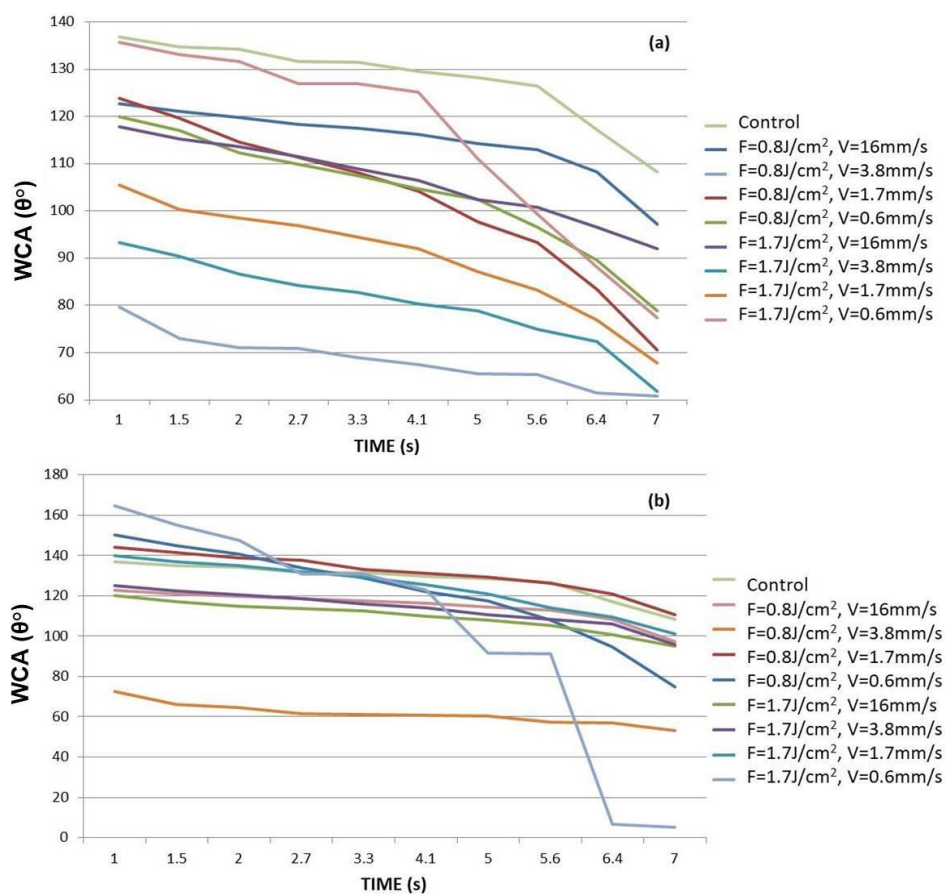

**Figure 9.** WCA evaluation of Fs modified PLA samples for a period of 7s – **(a)** the dH$_2$O drop applied along the direction of the grooves; **(b)** the dH$_2$O drop applied perpendicular to the direction of created grooves.

However, some interesting cases stand out from the others. In the case of water application along with the Fs stripes, the control PLA surface possess most hydrophobic features, which is not the case, when perpendicular dropping is performed - at the first seconds of application, part of the Fs treated PLA surfaces exhibit greater hydrophobicity than the non-treated PLA. However, at the 7$^{th}$ second of both types of water drop application, each laser modified surface changes its wettability properties to more hydrophilic. This could be explained by the Wenzel model (1936), which describes the homogeneous wetting on the rough surfaces at the stable equilibrium state - the minimum free energy state for the system, which the water drop demands [44]. The WCA gradually decreases with time except for the PLA sample treated with V=0.6 mm/s, F= 1.7 J/cm$^2$, where interesting bends (along and perpendicular the direction of the grooves - Figure 9 (a) and (b), respectively) in both WCA graphs are observed. These twists are more profoundly expressed when "perpendicular dropping" is performed. This could be explained by the model, established by Cassie and Baxter (1944), according to which a heterogeneous wetting of the surface is observed [45]. As can be seen from SEM images (Figure 6h) and 3D cross section images (Figure 7b) of V=0.6 mm/s, F= 1.7 J/cm$^2$ treated PLA samples, the formation of the porous periodical microstructured architecture of the PLA scaffold has a hierarchical origin and thus a preservation of air pockets in the volume of the material is possible, which prevent the initial uniform entry of the water droplet inside the laser created microstructure and could lead to the sudden jump-like decrease of the WCA values which forms the graph bends, observed in Figure 9. It was found out that superhydrophobic (θ >150°) or superhydrophilic (θ <20°) surfaces can be achieved by creation of grooves with V=0.6 mm/s, F= 1.7 J/cm$^2$ PLA, where θ=164.8° at the 1$^{st}$ second drops noticeably to θ=5° at the 7$^{th}$ second of application. This drastic jump-like change in WCA could be explained with the Cassie-Baxter model [45]. The WCA evaluation shows that by fine-tuning laser parameters, the surface wettability of PLA matrices can be precisely controlled in order to achieve cell-adhesive surfaces.

The chemical composition after laser treatment of the PLA cell matrices was also investigated. The results of the performed FTIR analysis are presented in Figure 10.

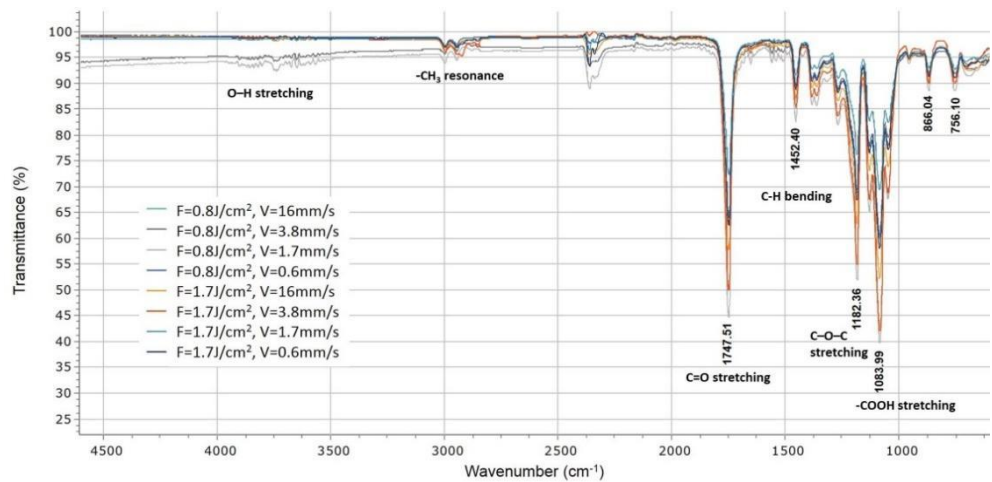

**Figure 10.** FTIR spectrum obtained for laser structured PLA surfaces treated by laser irradiation at F= 0.8 and 1.7 J/cm$^2$, V=16, 3.8, 1.7 and 0.6 mm/s.

As can be seen from the results obtained for different treatment conditions of the performed FTIR spectroscopy, in respect to control PLA [37], slight deviations in the intensity of the acquired peaks are monitored, after Fs-laser treatment is observed. All characteristic chemical groups for the PLA spectra are detected, deviations are observed at the intensity of the occurring peaks. The peak detected at 3750 cm$^{-1}$, corresponds to O–H bond stretching and is characteristic for the PLA. The resonance peak of the CH$_3$ group is detected at 2925 cm$^{-1}$. The maximums at 1747.51 cm$^{-1}$ and 1182 cm$^{-1}$ correspond to the C=O stretching and the C–O–C stretching of PLA molecules. The carbonyl peak at 1452.4 cm$^{-1}$ is well defined. The Carboxyl group was also detected at 1083.99 cm$^{-1}$ [46].

The obtained data via FTIR were additionally characterized by means of Principal Component Analysis (PCA) method, which is used as a chemometric methodology for analysis of data with high-dimensional sets. The basic aim of PCA is to gain a reduced set of principal components (PC), which provides explanation to the variability of obtained data. For visualizing clustering of multivariate data, the *Clustvis web tool* (Nucleic Acids Research, 43(W1):W566–W570, 2015) was employed - each treatment condition of the obtained FTIR spectra was assigned as a variable in PCA, which permits the discovery of changes of the samples characteristics in respect to physico-chemical properties.

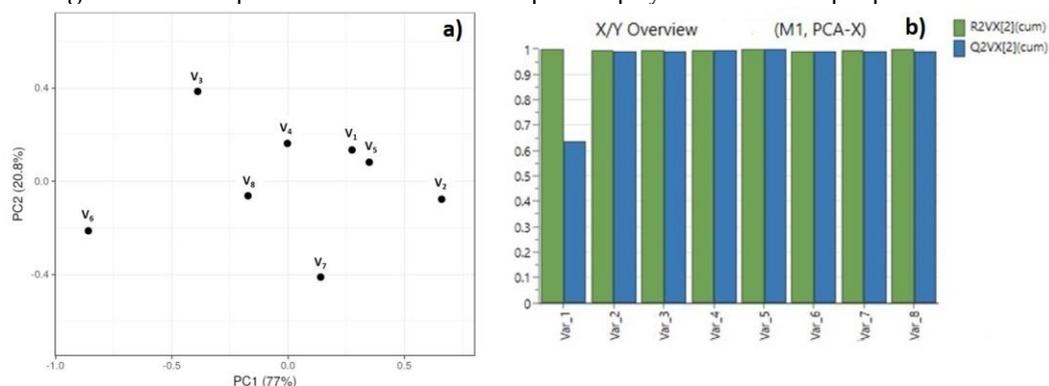

**Figure 11.** PCA scores scatter plots of FTIR spectra in the region from 600 - 4600 cm$^{-1}$; (a) V1 - F=1.663J/cm$^2$, V=1.7mm/s; V2 - F=1.663J/cm$^2$, V=3.8mm/s; V3 - F=1.663J/cm$^2$, V=16mm/s; V4 - F=0.831J/cm$^2$, V=6.0mm/s; V5 - F=0.831J/cm$^2$, V=1.7mm/s; V6 - F=0.831J/cm$^2$, V=16mm/s; V7 - F=0.831J/cm$^2$, V=3.8mm/s; V8 - F=1.663J/cm$^2$, V=0.6 mm/s; (b) The Summary of plot fit of the PCA model for component R2 and Q2.

The obtained variances are 77% and 20.8% for PC-1 and PC-2, respectively. The results for all sets of data for the first two principal components PC1 and PC2, jointly were calculated to be 97.8% of the data matrix variance. The scores scatter plot PC1 vs. PC2 (Figure 11a) showed peaks distribution and scattering along the PC1 axis. The V1, V3, V4, V5, are positioned on the positive side of PC2. In Figure 11 the tendency for the assembling of scores in a pattern going to negative values of PC2 is monitored. In our specific case the data refers to the same materials used for processing, thus in this concrete case we want to highlight the similarities between different processing regimes and applied laser parameters. Our analysis is directed towards estimation of possible contribution of separate laser treatment conditions to the deviation in chemical composition of the material, which under investigation of FTIR spectra remains almost equal, after different laser treatment conditions applied. This is clearly visible in the provided data on (Figure 11a). To summarize these findings, the analyzed data suggest that each treatment condition was well separated and scores showed the most significant variation lying along the PC2 axis. As a general tendency is observed, that the scores which appear in the negative part of the graph are related to the gentle regime of laser processing, associated with lower laser energy values applied, while the distribution of the components in the upper (positive part of the graph are related to laser treatment with increased laser fluence. The model prediction is shown on Figure 11b. Where are plotted cumulative values for R2 ( percent of variation of the training set – X with PCA - explained by the model) and Q2 (percent of variation of the training set – X with PCA – predicted by the model according to cross-validation). When R2 values are close to 1 then we can predict that our conditions are very close to a good model. Concerning Q2, it indicates how well the model predicts new data. A large value of Q2 (Q2 > 0.5) indicates good predictivity. In our specific case, we have for both values numbers approximately to 1, which defines a good approximation to the PCA model (Figure 11b).

Additionally, XPS analysis was performed at the two most interesting cases (adapted to achieve laser modification of PLA surface, optimized to achieve enhanced cell adhesion conditions) of laser parameters applied to PLA samples- F= 0.8 J/cm$^2$, V=3.8mm/s, and F= 1.7 J/cm$^2$, V=0.6 mm/s - Figure 12.

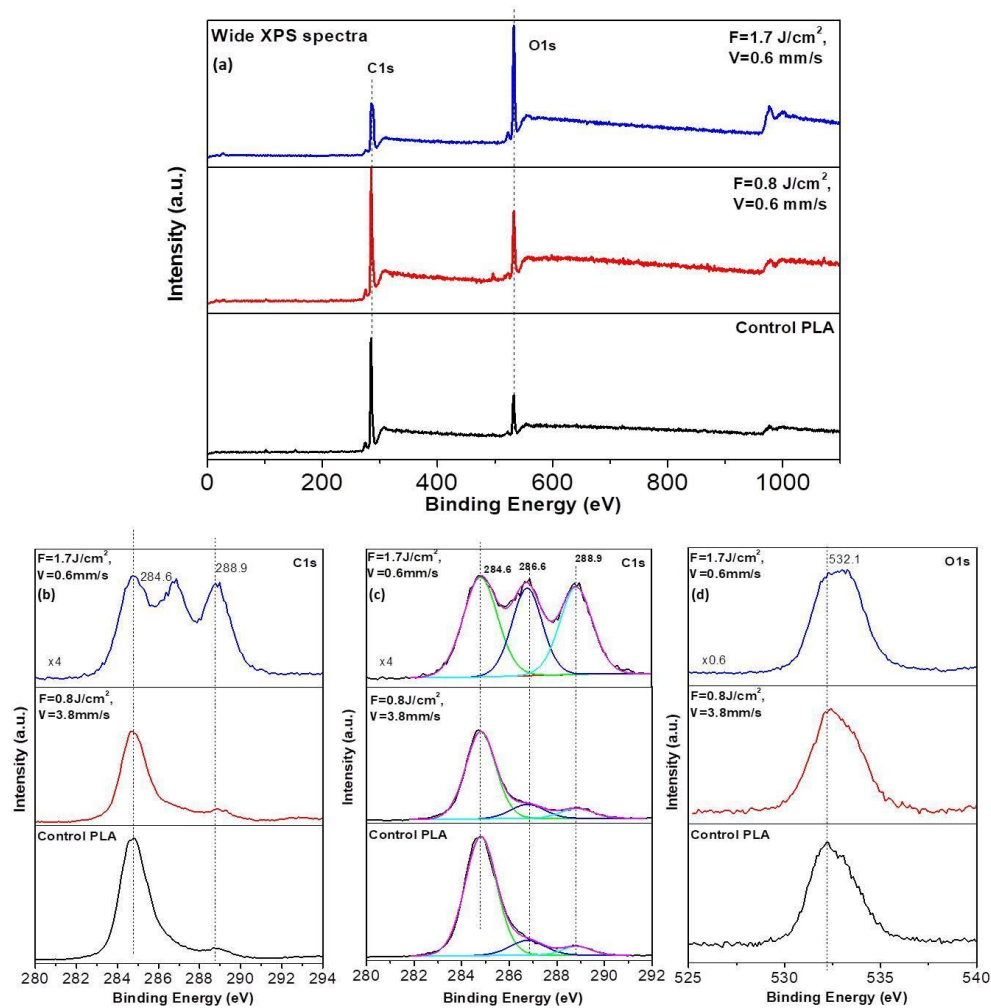

**Figure 12.** Comparison of XPS spectra of PLA surfaces treated by laser irradiation at F= 0.8 J/cm$^2$, V=3.8mm/s and F= 1.7 J/cm$^2$, V=0.6 mm/s; **(a)** Wide XPS Spectra; **(b)** C$_{1s}$ spectra; **(c)** C$_{1s}$ decomposed spectra; **(d)** O$_{1s}$ spectra.

Based on the PLA atomic composition, the C$_{1s}$, O$_{1s}$ regions were explored with XPS analysis performed as the biocompatible polymer is mainly composed of carbon and oxygen [47]. The wide XPS spectra show no deviation from the positions of the C$_{1s}$, O$_{1s}$ peaks obtained from the laser-treated zones of PLA, compared to control surfaces - Figure 12(a). C$_{1s}$ spectra (b) and C$_{1s}$ decomposed spectra (c). On the other hand, obtained spectra from treatments at F= 1.7 J/cm$^2$, V=0.6 mm/s significantly differ from control spectrum and that from F= 0.8 J/cm$^2$, V=3.8mm/s - Figure 12(b) and (c). In the case of F= 1.7 J/cm$^2$, V=0.6 mm/s fs treated matrices, three components at the C$_{1s}$ XPS spectra are observed – at 284.6, 286.6, and 288.9 eV. They can be assigned to C atoms in the (C–C) / (C–H), (C–O) and (O-C=O) bonds, respectively. In the case of the spectra detected from the control and the spectra obtained for irradiation conditions F= 0.8 J/cm$^2$, V=3.8mm/s, the second peak (286.6 eV) assigned to (C=O) bonds is almost not observed Figure 11(b), while in the case of F= 1.7 J/cm$^2$, V=0.6 mm/s this maximum is with significantly higher intensity, which could be related to an increase in the concentration of C–O and O-C=O bonds and corresponding incorporation of peroxyl, hydroxyl, ether or carbonyl groups on the PLA scaffold surface [47,48]. The significant increase in the third peak (288.9 eV) of C$_{1s}$ spectra suggests an increase in the molar concentration of C=O bonds and accompanying incorporation of carbonyl groups on PLA samples surface treated with F= 1.7 J/cm$^2$, V=0.6 mm/s. As can be seen from Figure 12(d), only a slight deviation in O$_{1s}$ XPS spectra is observed after laser treatment - the peak at 532.1 eV is attributed to oxygen atoms in (C–O) and (O-C=O) bonds.

In order to see if the degradation rate of the PLA scaffolds matches the speed of the new tissue regeneration at the side of implantation, a preliminary *in vitro* degradation study in PBS, 37°C of the PLA matrices chosen for cell seeding experiments, was carried out - **Group1 (G1)** – PLA F= 0.8 J/cm$^2$, V=3.8mm/s; **Group2 (G2)** – PLA F= 1.7 J/cm$^2$, V=0.551mm/s and **Group3 (G3)** – control PLA. The results of pH measurements obtained through a period of 8 weeks are summarized in Table 3 and Figure 13.

**Table 3.** pH change of PBS of **Group1 (G1)** – PLA F= 0.8 J/cm$^2$, V=3.8mm/s; **Group2 (G2)** – PLA F= 1.7 J/cm$^2$, V=0.6 mm/s and **Group3 (G3)** – control PLA during the 8 week period of *in vitro* degradation test.

| Week | Group 1 PBS pH | Group 2 PBS pH | Group 3 PBS pH |
|---|---|---|---|
| 0 | 7.2 | 7.2 | 7.2 |
| 1 | 7.2 | 7.2 | 7.2 |
| 2 | 7.2 | 7.2 | 7.2 |
| 3 | 7.2 | 7.2 | 7.2 |
| 4 | 7.2 | 7.2 | 7.2 |
| 5 | 7.2 | 7.2 | 7.2 |
| 6 | 7.1 | 6.9 | 7.1 |
| 7 | 6.9 | 6.7 | 6.9 |
| 8 | 6.5 | 6.3 | 6.6 |

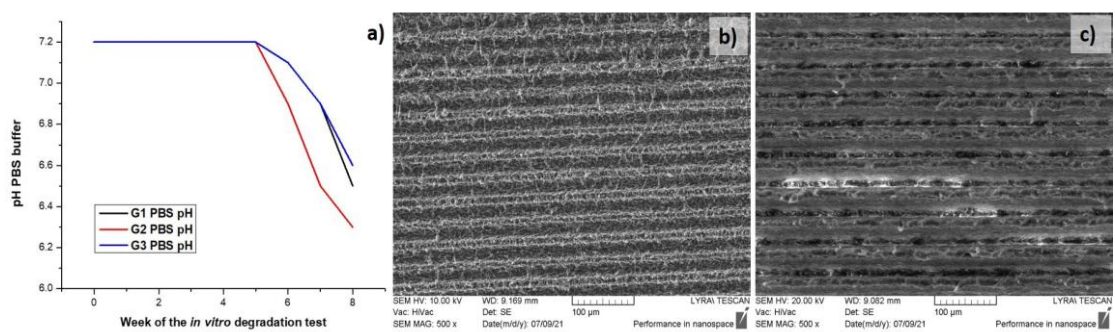

**Figure 13. (a)** Graphical presentation of pH change of PBS of **Group1, 2 and 3** over time; **(b)** SEM images of **G1, (c) G2** samples after the *in vitro* degradation test.

The graph presented on Figure 13 (a) shows that the PBS pH values of the three groups remained unchanged until the end of the fifth week (~ 7.2). At the sixth week a decrease in the pH value of the buffer of the **G1** (pH=6.9), **G2** (pH=6.7) and **G3** (pH=6.9) was observed, as it is more noticeable for **G2** samples. This tendency was kept until the end of the *in vitro* degradation test – the pH dropped with ~ 0.2 units respectively for each of the groups. The slight decrease of the pH values of the PBS saline could be attributed to PLA degradation process, which byproducts released (like lactic acid), make the medium more acidic. The SEM images taken for G1 and G2 after the period of 8 weeks buffer incubation showed the slight deviation (Figure 13 b and c) in the surface morphology of PLA as compared to Figure 6. The results show that Fs laser processing with higher fluence and lower scanning velocity could speed up/ accelerate the *in vitro* degradation rate of the polymer matrix, and thus assist in achieving appropriate speed, matching the specific tissue regeneration rate [49]. The group of Guo et al. for example, demonstrated that the synthesized PLA scaffold were degraded completely into harmless products in simulated body fluid (SBF) at a slow degradation rate and that the scaffolds weight loss could reach 80% after 8 months *in vitro* degradation, which matches the speed of new bone tissue regeneration in human body [50].

*3.2. Microbiology studies*

In order to see if the created laser induced microchannel complex frameworks, obtained by fs laser structuring type **G2** on PLA samples surface have potential antibacterial properties, microbiology studies with *S. aureus* were performed. The results of the conducted experiments with *S. aureus*, show that the adhesion of the bacteria is very high onto PLA samples treated with laser patterning type (**G2**) in comparison to the control (**G3**) scaffolds. The results from Fs laser surface structuring showed to promote bacterial adhesion, as a 5-fold increase in the number of adhered bacteria onto PLA-**G2** is observed in comparison to control PLA **G3** surface – Figure 14. The laser treatment creates "3D-cavities" into PLA **G2** samples, which favor bacteria adhesion.

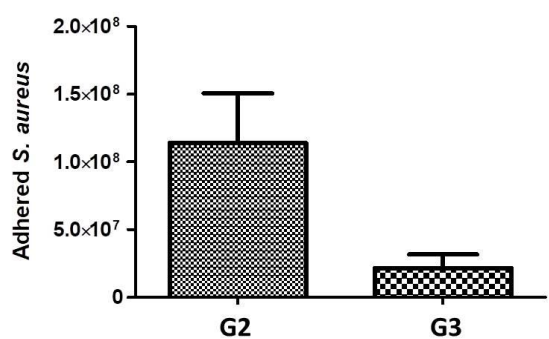

**Figure 14.** Number of adhered *S. aureus* bacteria on the grooved PLA samples surface (**G2** – PLA F= 1.7 J/cm$^2$, V=0.6 mm/s; **G3** – control PLA samples).

The group of Alves et al. for example showed experimentally that Fs laser created grooves, separated by distance of 8-12 μm and with channel width in the range 0.54 μm to 1.33 μm favor *S. aureus* adhesion. Further increase of surface roughness did not favor the bacteria adhesion and biofilm formation [51]. It is reported by other research groups that an intact bacteria biofilm occurred much faster when the surface roughness was greater than 0.8 μm as the contact area between the bacteria cell and the surface is increased [52,53]. On the other hand it is reported that such bacteria attach easier on hydrophilic surfaces (as **G2**) stronger, compared to hydrophobic, where the cells are easier to remove by body fluid circulation [54-56].

*3.3. Mesenchymal stem cell behavior*

After full characterization of the laser processed PLA matrices, two groups of Fs treated samples were chosen for preliminary cells experiments – **Group1 (G1)** – PLA F= 0.8 J/cm$^2$, V=3.8mm/s (in respect to the obtained R$_a$ and WCA values, which fall within the required conditions for MSCs adhesion and proliferation, according to the mentioned above literature survey [38-43]) and **Group2 (G2)** – PLA F= 1.7 J/cm$^2$, V=0.6 mm/s, which exhibits interesting wetting behavior and superhydrophilic nature due to its double hierarchical porosity. Each group consisted of 30 identical samples. As already mentioned, the MSCs proliferation experiments performed for a period of 14 days were repeated twice in order to confirm the results obtained. Parallel cell seeding was performed on control PLA samples **Group3 (G3)**. The results presented in Figure 14 (a) demonstrate that MSCs proliferation rate on **G1** is close to that of **G3,** while a slower MSCs proliferation rate was observed on **G2** matrices in comparison to the **G3**. Moreover, after the 14 days of culture on the three groups of PLA samples, MSCs were fixed, and fluorescently labeled with Phalloidin® (colors cytoskeleton in green) and DAPI (colors nuclei in blue). The observation of labeled cells on samples under fluorescence microscope confirmed the presence of MSCs on the samples after the 14 days of culture Figure 15 (b). From the figure it is observed, the Fs laser treatment influences MSCs morphology and organization – the cells cytoskeleton obtained an elongated shape simultaneously aligning along the grooves while on the PLA **G3** group cells keep a random organization. These preliminary results suggest that MSCs can adhere and orientate on both **G1** and **G2** types of patterning conditions and do not show differences between them in the context of cell viability.

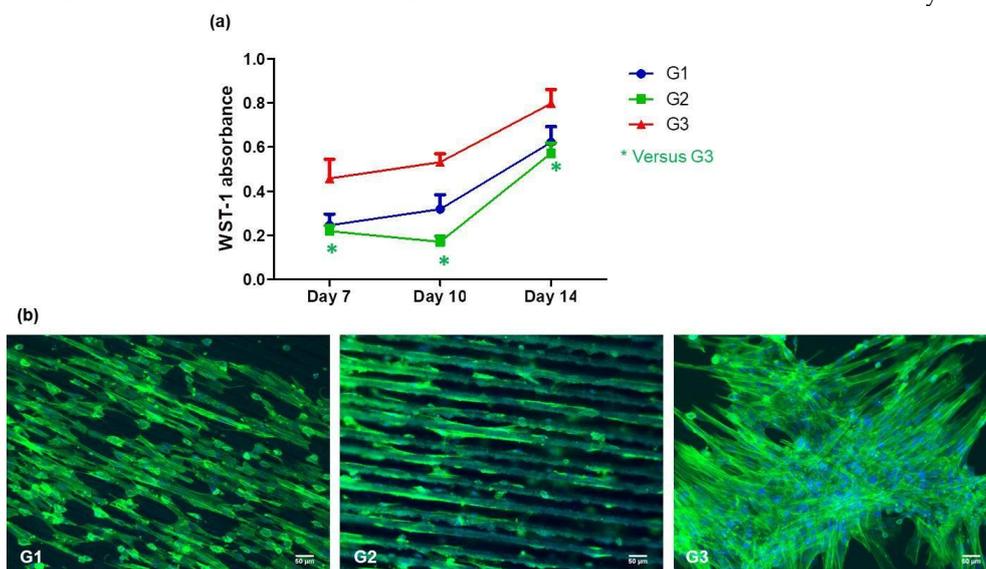

**Figure 15.** MSCs proliferation rate **(a)** and fluorescence images of morphology cellular cytoskeleton **(b)** on **G1**, **G2** and **G3** matrices. cellular cytoskeleton (in green) and nuclei (in blue). Scale bars=50 μm.

Femtosecond laser processing of polymer– based cell scaffolds has been extensively studied in last decades as an alternative contactless and chemically clean method for matrix surface structuring and functionalization, thus enhancing the biocompatibility and biomimetic properties of diverse polymer scaffolds [57-62]. The ultrashort time of interaction between the femtosecond laser pulse and the transparent biopolymers significantly reduces the size and depth of the heat affected zone, thus fs structuring of polymers like PLA, that have low glass-transition and melting temperatures, ($T_{g\,PLA}$=60°C) is possible to be achieved without triggering severe thermal side effects [35]. Not only surface micro and nano structuring, but also precise three-dimensional porous structures can be fabricated on top and in the interior of the processed biodegradable polymer [27]. Femtosecond laser systems enable the processing of transparent materials via nonlinear optical interaction in visible to near-IR light [63,64]. Ortiz et al. reported that the laser ablation of poly-L-lactic acid (PLLA) is governed by a multi-photon process in the three cases of interaction: with the fundamental (1064 nm), the second (532 nm) and the third harmonic generation wave (355 nm) of Nd: $YVO_4$ laser [65]. The group of Daskalova et al. for example achieved 3D porous structures composed of μm-sized porous foam on the surface of collagen, elastin and chitosan thin films by femtosecond laser structuring at λ=800 nm in which cells adhere, elongate, and orientate [66-68].

It has been shown that optimal cellular adhesion conditions in respect to protein absorption are achieved in the intermediate values of wettability – θ = 20–55° for fibroblasts and endothelial cells [42, 43]. Femtosecond laser processing of polymers leads to rupture of weak hydrogen bonds (C-H). The unsaturated C- bond on the main carbon chain reacts with oxygen and forms new carboxyl, ester, hydroxyl, carbonyl, etc. bonds, making the polymer-based cell scaffolds hydrophilic [30]. Enhanced surface roughness is very important for MSCs adhesion and further engineering of hard tissues like bone and teeth. For example, the group of Cordero et al. showed a better MC3T3-E1 pre-osteoblast cell alignment on laser micro patterns formed with line separation distances below 50μm [69]. Larger distances between structures prevent cells from developing focal adhesion plaques and bridge over the groove created on the polymer surface – the cells remain oriented along the length of the channels, but do not proliferate and do not express differentiation markers [70, 71]. The roughness of 0.9 - 1.53 μm ($R_a$) is reported as optimal for orientation of MSCs [38]. If the surface does not possess high roughness values, this would lead to insufficient formation of healthy tissue. Efficient implant to tissue bonding contacts could not be established, due to poor cell adhesion and tight fixation of the implant to the host tissue could fail [70]. On the other hand, roughness ($R_a$) values higher than 1μm are experimentally proven to be the minimum required for efficient tight bonding between the implant/tissue interfaces [39].

One of the most common reasons for implant rejection is bacterial infection [72]. The main efforts to prevent it are aimed at rupturing the bacteria biofilm, not allowing bacteria proliferation [72]. This could be achieved by inducing suitable surface irregularities, which could lead to stretching and disfiguration of bacteria, cell wall rupture, and death [73, 74]. The optimal goal to obtain biocompatible and anti-bacterial cell scaffold surfaces could be achieved by means of Fs-laser treatment. In its elaborate review, N. Sirdeshmukh and G. Dongre outline the biocompatibility and antibacterial analysis of laser surface textured patterns made by different research groups [75]. The scientists are pointing out that the right combination of micro and nano roughness could lead to the desired antibacterial effect, by means of LIPSS generation in combination with different nanostructures- nanopillars [76], spike nanotextured patterns [77], nanoripples [78], conic and spherical nanostructures [79], etc. - all mimicking topography of natural structured surfaces exhibiting antibacterial proper-ties [80]. On the other hand, as already mentioned, the generation of laser-induced periodic structures, on the polymer surface such as PLA could be a difficult task to achieve, which could be attributed to their low melting ($T_m$) and glass transition ($T_{g\ PLA}$=60°C) temperature [35]. The importance of the nanostructures could be explained by the attachment point theory according to which, if the surface available is smaller or near the size of the bacteria, not enough attachment strength can be achieved, and bacteria can not adhere effectively (*S. aureus* - d=0.5÷1.5 μm) [81,82]. The group of Guenther is reporting that laser-induced pillar-like patterns with 0.5 μm spatial period significantly decrease *S. aureus* attachment on structured polymer surfaces [83].

We demonstrated experimentally that by simply tuning the Fs laser parameters applied – fluence and scanning velocity, precise control of PLA scaffold surface properties, like roughness, porosity, and wettability, without changing its elemental composition and chemical structure, could be achieved. The results presented clearly show no cell cytotoxicity and a tendency of cell adhesion and elongation along the grooves, which could orientate them in the desired direction of the tissue regeneration. The results from microbiology studies with *S. aureus* demonstrated that such surface modifications could find application as "bacteria traps" in clinical practice, "luring" microorganisms as an optimal environment, leaving in that way the desired surface (bio-sensors, biochips, drug carriers, prostheses for medical implants, etc.) clean of them. The preferential alignment of *S. aureus* to the periodic structures, created by the Fs laser on the PLA surface, demonstrates the microorganism's attempt to improve their surface retention by maximizing their attachment points with the laser structured PLA scaffold [84-86]. Additional nanostructuring of the achieved laser-induced micro-channel complex frameworks is therefore expected to reduce bacteria attachment by limiting the number of contact adhesion points with the PLA implant surface. Further optimization of Fs laser parameters used for surface structuring in the context of anti-bacterial effect and investigation of the results obtained should be made in order for bio-compatible and antibacterial PLA cell scaffolds and bio-interfaces to be achieved.

## 5. Conclusions

The experimental results obtained in the current study demonstrate that Fs micropatterning could improve biological properties of PLA cell scaffolds in the context of directional orientation, and biocompatibility. By controlling the laser parameters (laser fluence and scanning velocity) PLA bioactive surfaces can be achieved via a contactless and non-destructive laser modification process, which leads to precise control of PLA cell-matrix surface properties like roughness, wettability, and porosity.

PLA sheets were patterned in a form of rows using fs-laser emitting laser radiation with 150-fs pulse duration. In order to achieve specific patterning the optimal laser fluence and scanning velocity were determined as F= 0.8 J/cm$^2$, V=3.8mm/s, and PLA F= 1.7 J/cm$^2$, V=0.6mm/s ,respectively. At these applied laser parameters The PLA polymer matrix formed a hierarchical structure. The examination of degradation properties before and after laser patterning, showed degradation of the processed and control PLA samples after the 6th week of the *in vitro* degradation test, with emphasis on the samples modified with increased laser fluence. These findings are explained by the decrease in crystallinity related to the occurrence of surface melting, in a high fluence regime. The obtained results of the wettability test at different laser processing parameters demonstrate that some laser-patterned designs have the potential to create stable hydrophilic properties, while in some specific cases, obtaining hierarchical structures, a sudden drop of contact angle values was observed due to air pockets formation. Principal component analysis was employed to distinguish the main origin of variance in the FTIR spectra of PLA samples obtained under diverse treatment conditions. PCA permits the samples to be analyzed in terms of their similarities, in accordance with applied Fs laser processing, with respect to the acquired FTIR spectra.

The Fs laser treatment influences MSCs morphology and orientation along the grooves formed by the laser, compared to the chaotic spreading of cells, cultured on control Fs non-treated PLA scaffolds, without affecting cells viability. Generation of hierarchical structure on the PLA surface was monitored, and creation of ideal cytocompatible structure was obtained.

**Author Contributions:** Conceptualization, A.D.; methodology, A.D.; writing, A.D.; original draft preparation, A.D.; project administration, A.D.; writing, L.A.; original draft preparation, L.A.; investigation L.A.; investigation, E.F; investigation, D.A.; investigation, R.M.; investigation, X.C.; methodology, H.K.; visualization, H.K.; methodology, M.D.; visualization, M.D.; investigation, J.C.; visualization, J.C.; software, A.T.; supervision, I.B. All authors have read and agreed to the published version of the manuscript.

**Funding:** This work was supported by: EUROPEAN UNION'S H2020 research and innovation program under the Marie Sklodowska-Curie Grant Agreement AIMed No. 861138; BULGARIAN NATIONAL SCIENCE FUND (NSF) under grant number No. KP-06-H48/6 (2020–2023), „Development of hybrid functional micro/nanoporous biomaterial scaffolds by ultra-fast laser modification"; and H2020 FET Open METAFAST Grant Agreement No. 899673.

**Conflicts of Interest:** The authors declare no conflict of interest.